# Electric Field Effect in Atomically Thin Carbon Films

K.S. Novoselov[1], A.K. Geim[1], S.V. Morozov[2], D. Jiang[1], Y. Zhang[1], S.V. Dubonos[2], I.V. Grigorieva[1], A.A. Firsov[2]

[1]Department of Physics, University of Manchester, M13 9PL, Manchester, UK

[2]Institute for Microelectronics Technology, 142432 Chernogolovka, Russia

*We describe monocrystalline graphitic films, which are just a few atoms thick but nonetheless stable under ambient conditions, metallic and of remarkably high quality. The films are found to be a two-dimensional semimetal with a tiny overlap between valence and conductance bands and to exhibit a strong ambipolar electric-field effect such that electrons and holes in concentrations up to $10^{13} cm^{-2}$ and with room-temperature mobilities $\approx$10,000 $cm^2/Vs$ can be induced by applying gate voltage.*

One-sentence summary: We report a naturally-occurring two-dimensional material – graphene that can be viewed as a gigantic flat fullerene molecule, – describe its electronic properties and demonstrate all-metallic field-effect transistor, which uniquely exhibits ballistic transport at submicron distances even at room temperature.

The ability to control electronic properties of a material by externally applied voltage is at the heart of modern electronics. In many cases, it is the so-called electric field effect that allows one to vary the carrier concentration in a semiconductor device and, consequently, change an electric current through it. As the semiconductor industry is nearing the limits of performance improvements for the current technologies dominated by silicon, there is a constant search for new, non-traditional materials whose properties can be controlled by the electric field. The most notable examples of such materials developed recently are organic conductors [1] and carbon nanotubes [2]. It has long been particularly tempting to extend the use of the field effect to metals (e.g., to develop all-metallic transistors that could be scaled down to much smaller sizes and also have the potential to consume less energy and operate at higher frequencies than traditional semiconducting devices [3]). However, this would require atomically thin metal films because the electric field is screened at extremely short distances (<1 nm) and bulk carrier concentrations in metals are large compared to the surface charge that can be induced by the field effect. Films so thin are thermodynamically unstable and become discontinuous already at thicknesses of many nm; so far, this has proved to be an insurmountable obstacle to metallic electronics and no metal or semimetal has been shown to exhibit any notable (>1%) field effect [4].

Here we report the electric field effect in a naturally occurring two-dimensional (2D) meterial that we refer to as few-layer graphene (FLG). Graphene is the name given to a single layer of carbon atoms densely packed into a benzene-ring structure. This hypothetical material is widely used to describe properties of many carbon-based materials, including graphite, large fullerenes, nanotubes, etc. (e.g., carbon nanotubes are usually thought of as graphene sheets rolled up into nm-sized cylinders) [5-7]. Planar graphene itself has so far been presumed not to exist in the free state, being rather unstable with respect to the formation of curved structures such as soot, fullerenes and nanotubes [5-14]. Contrary to the common belief, we have been able to prepare graphitic sheets of thicknesses down to a few atomic layers, including single layer graphene, and succeeded in making devices from them and studying their electronic properties. Despite being atomically thin, the films remain of surprisingly high quality so that 2D electronic transport is ballistic at submicron distances. This is truly remarkable, as no other film of similar thickness is known to be even poorly metallic or continuous under ambient conditions. Using FLG, we demonstrate a metallic field-effect transistor, in which the conducting channel can be switched between 2D electron and hole gases by changing gate voltage.

The reported graphene films were made by mechanical exfoliation (repeated peeling) of small mesas of highly-oriented pyrolytic graphite as described in the supporting online material [15]. This approach was found to be highly reliable and allowed us to prepare FLG films up to 10 μm in size. Thicker films ($d \geq 3$nm) were up to a hundred microns across and visible by the naked eye. Figure 1 shows examples of the prepared films, including single-layer graphene (see also [15]). To study their electronic properties, the films were processed into multi-terminal Hall bar devices placed on top of an oxidized Si substrate so that gate voltage $V_g$ could be applied. We have studied more than 60 devices with $d$ <10 nm. In this report, we focus on electronic properties of our thinnest (FLG) devices, which contained just 1, 2 or 3 atomic layers [15]. All FLG devices exhibited essentially identical electronic properties characteristic for a 2D semimetal, which at the same time were drastically different from a more complex (2D plus 3D) behavior observed for thicker, multilayer graphene [15] as well as from the properties of 3D graphite.

Figure 2 shows typical dependences of resistivity ρ and the Hall coefficient $R_H$ in FLG on gate voltage $V_g$. One can see that ρ exhibits a sharp peak to a value of several kΩ and decays to ≈100Ω at high $V_g$. In terms of conductivity σ =1/ρ, it increases linearly with $V_g$ on both sides of the resistivity peak (Fig. 2B). At the same $V_g$ where ρ has its peak, $R_H$ exhibits a sharp reversal of its sign (Fig. 2C). The observed behavior resembles the ambipolar field effect in semiconductors but there is no zero-conductance region associated with the Fermi level being pinned inside the band gap.

Our measurements can be explained quantitatively by a model of a 2D metal with a small overlap δε between conductance and valence bands [15]. The gate voltage induces a surface charge density $n = \varepsilon_0 \varepsilon V_g/te$ and, accordingly, shifts the position of the Fermi energy $\varepsilon_F$. Here, $\varepsilon_0$ and $\varepsilon$ are permittivities of free space and SiO$_2$, respectively, $e$ is the electron charge, $t$ the thickness of our SiO$_2$ layer (300 nm). For typical $V_g$ =100V, the formula yields $n \approx 7.2 \cdot 10^{12}$ cm$^{-2}$. The electric-field doping transforms the shallow-overlap semimetal into either completely electron or completely hole conductor through a mixed state where both electrons and holes are present (see Fig. 2). The three regions of electric-field doping are clearly seen on both experimental and theoretical curves. For the regions with only electrons or holes left, $R_H$ decreases with increasing the carrier concentration in the usual way, as $1/ne$. The resistivity also follows the standard dependence $\rho^{-1} = \sigma = ne\mu$. In the mixed state, σ changes little with $V_g$, indicating the substitution of one type of carriers with another, while the Hall coefficient reverses its sign, reflecting the fact that $R_H$ is proportional to the difference between electron and hole concentrations.

Without electric-field doping (at zero $V_g$), FLG was found to be a hole metal, which is seen as a shift of the peak in ρ to large positive $V_g$. However, this shift could be due to an unintentional doping of the films by absorbed gas molecules [16,17]. Indeed, we found that it was possible to change the position of the peak by annealing our devices in vacuum, which usually resulted in shifting of the peak close to zero voltages. Exposure of the annealed films to either water vapor or NH$_3$ led to their $p$- and $n$-doping, respectively. Therefore, we believe that intrinsic FLG is a mixed-carrier semimetal.

Carrier mobilities in FLG were determined from field-effect and magnetoresistance measurements as μ = σ($V_g$)/$en$($V_g$) and μ = $R_H$/ρ, respectively. In both cases, we obtained the same values of μ, which varied from sample to sample between 3,000 and 10,000 cm$^2$/V·s. The mobilities were practically independent of temperature $T$, indicating that even being so high they were still limited by scattering on defects. For μ ≈10,000 cm$^2$/V·s and our typical $n \approx 5 \cdot 10^{12}$ cm$^{-2}$, the mean free path is ≈0.4 μm, which is highly surprising given that the 2D gas is at most a few Å away from the interfaces. However, our findings are in agreement with equally high μ observed for intercalated graphite [5], where charged dopants are located next to graphene sheets. Carbon nanotubes also exhibit very high μ but this is commonly attributed to the suppression of scattering in the 1D case. Note that for multilayer graphene, we observed even higher mobilities, up to ≈15,000 cm$^2$/V·s at 300K and ≈60,000 cm$^2$/V·s at 4K.

Remarkably, despite being essentially gigantic fullerene molecules and unprotected from the environment, FLG films exhibit pronounced Shubnikov-de Haas (ShdH) oscillations in both longitudinal resistivity $\rho_{xx}$ and Hall resistivity $\rho_{xy}$ (Fig. 3). This serves as yet another indicator of the quality and homogeneity of the experimental system. Studies of ShdH oscillations confirmed that electronic transport in FLG was strictly 2D, as one could reasonably expect, and allowed us to fully characterize its charge carriers. First, we carried out the standard test and measured ShdH oscillations for various angles θ between the magnetic field and the graphene films. The oscillations depended only on the perpendicular component of the magnetic field $B \cdot \cos\theta$, as expected for a 2D system. More importantly, however, we found a linear dependence of ShdH oscillations' frequencies $B_F$ on $V_g$ (Fig. 3), yielding that the Fermi energies $\varepsilon_F$ of holes and electrons were proportional to their concentrations $n$. This dependence is qualitatively different from the 3D dependence $\varepsilon_F \propto n^{2/3}$ and unequivocally proves the 2D nature of charge carriers in FLG. Further analysis [15] of ShdH oscillations showed that only a single spatially-quantized 2D subband was occupied up to the maximum concentrations achieved in our experiments (≈3·10$^{13}$cm$^{-2}$). It could be populated either by electrons with mass $m_e \approx 0.06m_0$ located in two equivalent valleys or by light and heavy holes with masses ≈0.03$m_0$ and ≈0.1$m_0$ and the double-valley degeneracy ($m_0$ is the free electron mass). These properties were found to be the same for all studied FLG films and are notably different from the electronic structure of both multilayer graphene [15] and bulk graphite [5-7]. Note that graphene is expected to have the linear energy dispersion and carriers with zero mass, and the reason why the observed behavior is so well described by the simplest free-electron model remains to be understood. [15]

We have also determined the band overlap $\delta\varepsilon$ in FLG, which varied from 4 to 20meV for different samples, presumably indicating a different number of graphene layers involved. To this end, we first used a peak value $\rho_m$ of resistivity to calculate typical carrier concentrations in the mixed state, $n_0$ (e.g., at low $T$ for the sample in Fig. 2A-C with $\mu \approx 4,000$ cm$^2$/V and $\rho_m \approx 8$k$\Omega$, $n_0$ was $\approx 2\cdot 10^{11}$cm$^{-2}$). Then, $\delta\varepsilon$ can be estimated as $n_0/D$ where $D = 2m_e/\pi\hbar^2$ is the 2D density of electron states. For the discussed sample, $\delta\varepsilon$ is $\approx 4$meV, i.e. much smaller than the overlap in 3D graphite ($\approx 40$meV). Alternatively, $\delta\varepsilon$ could be calculated from the temperature dependence of $n_0$, as it characterizes relative contributions of intrinsic and thermally excited carriers. For a 2D semimetal, $n_0(T)$ varies as $n_0(0K)\cdot f\cdot \ln[1+\exp(1/f)]$ where $f = 2k_B T/\delta\varepsilon$, and Fig. 2D shows the best fit to this dependence, which yields $\delta\varepsilon \approx 6$meV. Different FLG devices were found to exhibit the ratio of $n_0(300K)/n_0(0)$ between 2.5 and 7, whereas for multilayer graphene it was only $\approx 1.5$ (Fig. 2D). This clearly shows that $\delta\varepsilon$ decreases with decreasing number of graphene layers. The observed major reduction of $\delta\varepsilon$ is in agreement with the fact that single-layer graphene is in theory a zero-gap semiconductor [5,18].

As concerns the metallic transistor, graphene is not only the first but also probably the best possible metal for such applications. In addition to the scalability to true nm sizes envisaged for metallic transistors, graphene also offers ballistic transport, linear I-V characteristics and huge sustainable currents (>$10^8$A/cm$^2$) [15]. Graphene transistors show a rather modest on-off resistance ratio (less than $\approx 30$ at 300K; limited because of thermally excited carriers) but this is a fundamental limitation for any material without a band gap exceeding $k_B T$. Nevertheless, such on-off ratios are considered sufficient for logic circuits [19] and it is feasible to increase the ratio further by, e.g., using p-n junctions, local gates [3] or the point contact geometry. However, by analogy to carbon nanotubes [2], it could be other, non-transistor applications of this unique molecular material, which ultimately may prove to be the most exciting.

## LIST OF REFERENCES

1. C.D. Dimitrakopoulos, D.J. Mascaro, *IBM J. Res. & Dev.* **45**, 11 (2001).
2. R.H. Baughman, A.A. Zakhidov, W.A. de Heer, *Science* **297**, 787 (2002).
3. See, e.g., S.V. Rotkin, K. Hess, *Appl. Phys. Lett.* **84**, 3139 (2004).
4. A.V. Butenko *et al*, *J. Appl. Phys.* **88**, 2634 (2000).
5. M.S. Dresselhaus, G. Dresselhaus, *Adv. Phys.* **51**, 1 (2002).
6. I.L. Spain, in *Chemistry and physics of carbon*, edited by P.L. Walker & P.A. Thrower **16**, 119 (Marcel Dekker Inc, New York, 1981).
7. O.A. Shenderova, V.V.Zhirnov, D.W. Brenner, *Crit. Rev. Sol. State Mat. Sci.* **27**, 227 (2002).
8. A. Krishnan *et al*, *Nature* **388**, 451 (1997).
9. E. Dujardin *et al*, *Appl. Phys. Lett.* **79**, 2474 (2001).
10. H. Shioyama, *J. Mat. Sci. Lett.* **20**, 499 (2001).
11. A.M. Affoune *et al*, *Chem. Phys. Lett.* **348**, 17 (2001).
12. K. Harigaya *et al*, *J. Phys. Cond. Mat.* **14**, L605 (2002).
13. T.A. Land *et al*, *Sur. Sci.* **264**, 261 (1992).
14. The closest known analogues to FLG were nanographene (nm-sized patches of graphene on top of HOPG) [11,12], carbon films grown on chemically binding metal surfaces [13] and mesoscopic graphitic disks with thickness down to $\approx 60$ graphene layers [8,9].
15. See supporting material on *Science Online* for details.
16. J. Kong *et al*, *Science* **287**, 622 (2000).
17. M. Krüger *et al*, *New J. Phys.* **5**, 138 (2003).
18. Non-zero values of $\delta\varepsilon$ found experimentally could also be due to inhomogeneous doping, which could smear the zero-gap state over a small range of $V_g$ and lead to finite apparent $\delta\varepsilon$.
19. M. R. Stan *et al*, *Proc. IEEE* **91**, 1940 (2003).

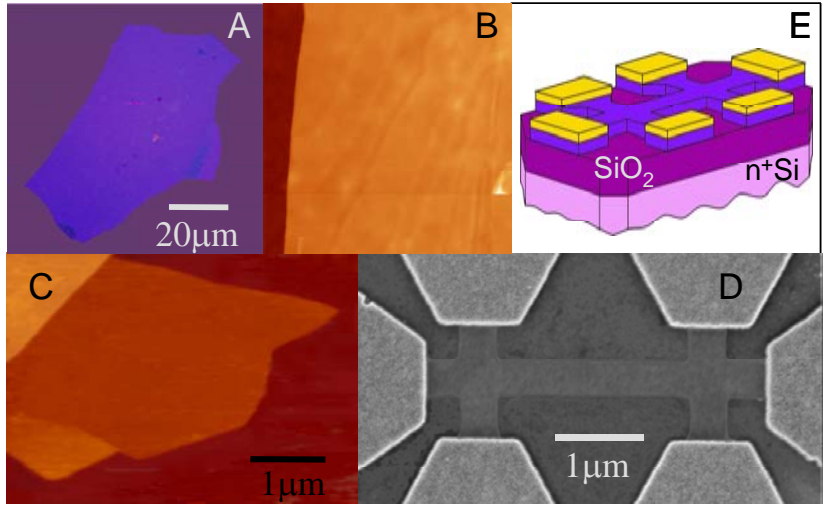

Figure 1. Graphene films. (**A**) Photograph (in normal white light) of a relatively large multilayer graphene flake with thickness ≈3nm on top of an oxidized Si wafer. (**B**) AFM image of 2x2 µm² area of this flake near its edge. Colors: dark brown is $SiO_2$ surface; orange corresponds to 3nm height above the $SiO_2$ surface. (**C**) AFM image of single-layer graphene. Colors: dark brown – $SiO_2$ surface; brown-red (central area) – 0.8nm height; yellow-brown (bottom-left) – 1.2nm; orange (top-left) – 2.5nm. Notice the folded part of the film near the bottom, which exhibits a differential height of ≈0.4nm. For details of AFM imaging of single-layer graphene, see [15]. (**D**) SEM micrograph of one of our experimental devices prepared from FLG, and (**E**) their schematic view.

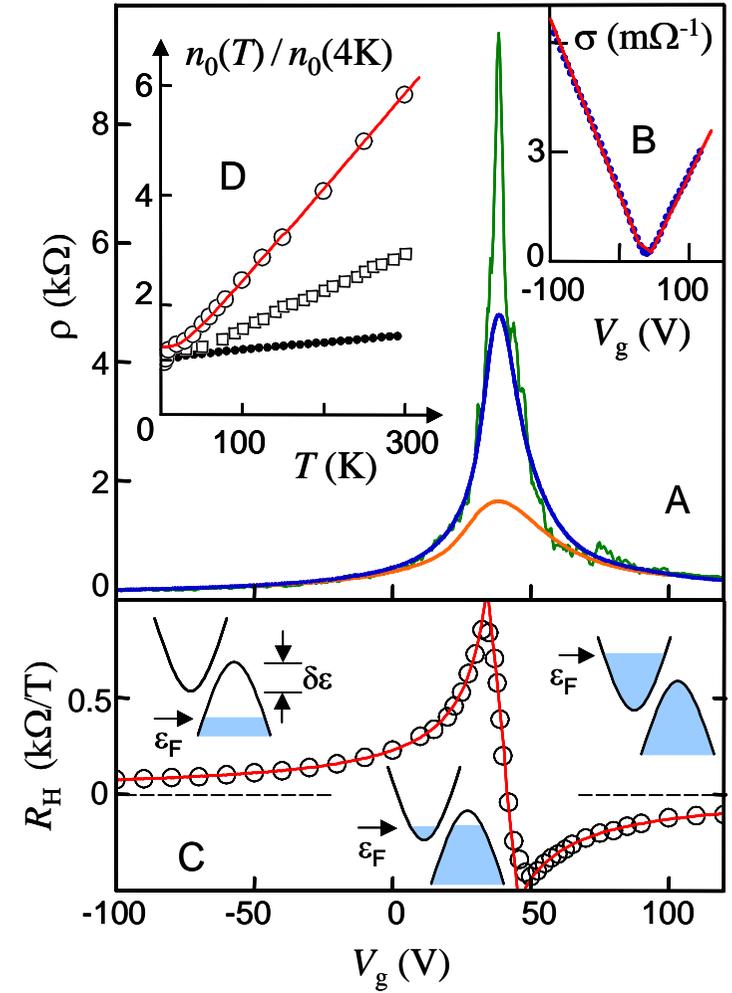

Figure 2. Field effect in few-layer graphene. (**A**) Typical dependences of graphene's resistivity ρ on gate voltage for different temperatures ($T$ =5, 70 and 300K for top to bottom curves, respectively). (**B**) Example of changes in the film's conductivity σ=1/ρ($V_g$) obtained by inverting the 70K curve (dots). (**C**) Hall coefficient $R_H$ vs $V_g$ for the same film. (**D**) Temperature dependence of carrier concentration $n_0$ in the mixed state for the film in (A) (open circles), a thicker FLG film (squares) and multilayer graphene ($d$ ≈5nm; solid circles). Red curves in (B) to (D) are the dependences calculated from our model of a 2D semimetal illustrated by insets in (B).

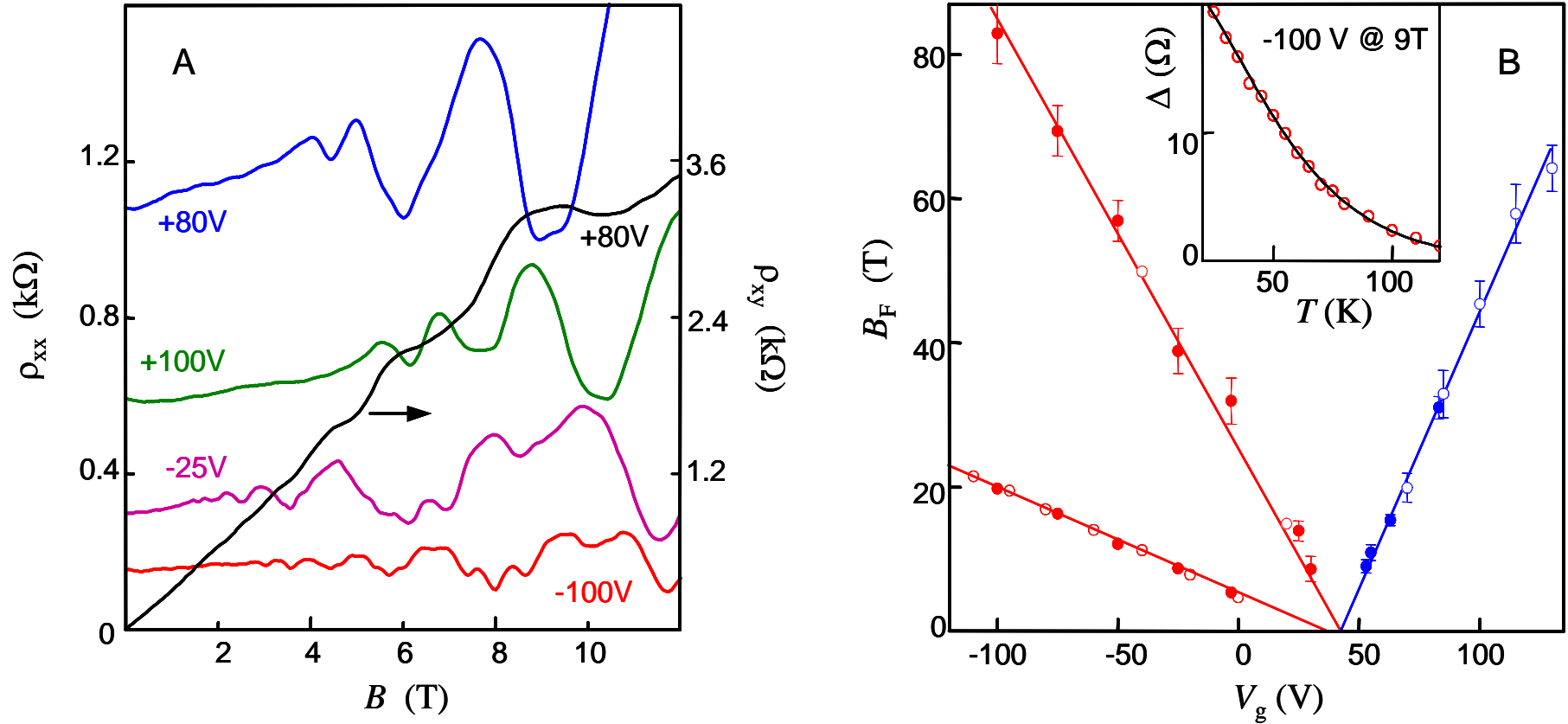

Figure 3. (A) Examples of Shubnikov-de Haas oscillations for one of our FLG devices for different gate voltages; $T=3K$ and $B$ is the magnetic field. As the black curve shows, we often observed pronounced plateau-like features in $\rho_{xy}$ at values close to $(h/4e^2)/\nu$ (in this case, $\varepsilon_F$ matches the Landau level with $\nu=2$ at around 9T). Such not-fully developed Hall plateaus are usually seen as an early indication of the quantum Hall effect in the situations where $\rho_{xx}$ does not yet reach the zero-resistance state. (B) Dependence of the frequency of ShdH oscillations $B_F$ on gate voltage. Solid and open symbols are for samples with $\delta\varepsilon \approx 6$ and 20 meV, respectively. Solid lines are guides to the eye. The linear dependence $B_F \propto V_g$ proves a constant (i.e., 2D) density of states [15]. The observed slopes (solid lines) account for the entire external charge $n$ induced by gate voltage, confirming that there are no other types of carriers and yielding the double-valley degeneracy for both electrons and holes [15]. The inset shows an example of the temperature dependence of amplitude $\Delta$ of ShdH oscillations (symbols), which is fitted by the standard dependence $T/\sinh(2\pi^2 k_B T/\hbar\omega_c)$ where $\omega_c$ is their cyclotron frequency. The fit (solid curve) yields light holes' mass of $0.03 m_0$.